\documentclass[twocolumn, amssymb, amsmath, preprintnumbers]{revtex4}

\usepackage{hyperref}
\usepackage{amsmath}
\usepackage{latexsym}
\usepackage{amssymb}
\usepackage[all]{xy}
\usepackage{graphicx}
\usepackage{epsfig}
\usepackage{psfrag}
\usepackage{textcomp}

\begin{document}  

\preprint{ITFA 2009-03, arXiv:0902.0002v1}
\title{A Black Hole Levitron$^{TM}$}
\author{Xerxes D. Arsiwalla}
\email{X.D.Arsiwalla@uva.nl}
\author{Erik P. Verlinde}
\email{E.P.Verlinde@uva.nl}
\affiliation{ Institute for Theoretical Physics, University of Amsterdam\\
              Valckenierstraat 65, 1018 XE Amsterdam, The Netherlands }

\bigskip

\date{February 2009}

\begin{abstract}
We study the problem of spatially stabilising four dimensional extremal black holes in background electric/magnetic fields. Whilst looking for stationary stable solutions describing black holes kept in external fields we find that taking a continuum limit of Denef et al's multi-center solutions provides a supergravity description of such backgrounds within which a black hole can be trapped in a given volume. This is realised by levitating a black hole over a magnetic dipole base. We comment on how such a construction resembles a mechanical Levitron$^{TM}$.
\end{abstract}

\maketitle 

\section{Introduction}
Being motivated by on-going interest in questions concerning black hole production; in this note we address a curiosity regarding how one could go about stabilising such a black hole using external fields, thus leading to a black hole analog of a particle-trapor rather as we shall see that of a    Levitron$^{TM}$.  However  unlike the more   familiar subatomic particle traps or even Millikan's famous oil drop experiment \cite{M}, the effects of general relativity give rise to interesting new features. We shall describe how this idea can in fact be materialised by writing down solutions for black holes levitating in electromagnetic as well as constant gravitational fields.

For the purpose of this note we consider four dimensional extremal black holes with electric/magnetic charges $q$ and $p$ respectively.  Extremal black holes are BPS solutions to four dimensional supergravity. The most general metric ansatz consistent with supersymmetry can be written as ( \cite{D1}, \cite{D2}, \cite{D3} )
\begin{eqnarray}
ds^2 &=& - \frac{\pi}{S(\vec{x})} (dt + \omega_i dx^i)^2  +  \frac{ S(\vec{x}) }{ \pi }  dx^i dx^i    \nonumber  \\
\mbox{with} \quad  S(\vec{x})/\pi &=& {\cal P}^2(\vec{x}) + {\cal Q}^2(\vec{x})      \nonumber  \\
\mbox{and} \quad  {\cal A} &=&  2 \pi  {\cal Q} (\vec{x})  \left( dt + \omega_i dx^i  \right)  +  \Theta
\label{1}
\end{eqnarray} 
is the four dimensional gauge field. ${\cal P}(\vec{x})$, ${\cal Q}(\vec{x})$ are harmonic functions associated to charges $p$ and $q$ respectively.  $\Theta$  is the Dirac part of the vector potential satisfying $d \Theta = \ast d {\cal P}(\vec{x})$ with the Hodge star $\ast$ defined on ${\mathbb R}^3$.   For  a  single   spherically symmetric black hole in vacuum, it holds that $\vec{\omega} = 0$.  However for our considerations, we shall be looking for solutions when the black hole is placed in external electric and magnetic fields. There is now a non-zero Poynting vector corresponding to a rotating geometry.  We first look for levitating  solutions in constant background fields. It turns out these are inadequate for stabilisation in all three directions. Then we look for   more non-trivial  backgrounds obtained using a continuum limit of Denef et al's  \cite{D1}, \cite{D2}, \cite{D3}  multi-center solutions and find that turning on dipole fields achieves the desired result.

\section{Black hole levitation in constant external fields}
Given the metric ansatz in eq.(\ref{1}), we begin by looking for stationary  solutions of a black hole placed in constant electric, magnetic and gravitational fields. In order to achieve this we have to specify explicit harmonic functions describing this configuration, then compute the off-diagonal elements $\vec{\omega}$ and solve the associated integrability equations. We claim that the desired   harmonic functions describing this configuration are
\begin{equation}
{\cal P}(\vec{x}) = u + \frac{p}{|\vec{x} - \vec{l}|}  + Bz    \qquad
  {\cal Q}(\vec{x}) = v + \frac{q}{|\vec{x} - \vec{l}|}  + Ez
\label{2}
\end{equation} 
where $B$ and $E$ are constant magnetic respectively electric fields oriented along the z-direction and $z$ denotes the z-coordinate. $\vec{l}$ marks the position of the black hole's horizon,  which we determine via integrability conditions. $u$, $v$ are constants. In principle, we can absorb $u$ and $v$ via a shift in the z-coordinate. This point will be made clear when we solve for $\vec{l}$.  The $Bz$ and $Ez$ in eq.(\ref{2}) are linear terms that satisfy Laplace's equation and can be recognised as the usual electro/magneto-static potentials associated to constant fields. Note that extremality implies the above linear terms also source constant gravitational fields. 

A nice way to motivate the expressions for ${\cal P}(\vec{x})$ and ${\cal Q}(\vec{x})$  is to extract them via a special limit of Denef et al's multi-center solutions  \cite{D1}, \cite{D2}, \cite{D3}. More specifically, let us consider the   two-center solution.  This is a regular BPS solution of four dimensional ${\cal N} = 2$ supergravity. It is stationary but non-static and hence caries an intrinsic angular momentum.  Moreover the black holes comprising this bound state possess mutually non-local charges.  Let us denote the corresponding two charge vectors as ${\mathbf    \Gamma} = (p,q)$ and $\mathbf{\tilde{\Gamma}}   = (\tilde{p},\tilde{q})$.  The idea is now to carry the  charge $\mathbf{\tilde{\Gamma}}$  all the way to infinity while scaling $(\tilde{p},\tilde{q})$  and the radial coordinate of the charges in such a way that the magnitudes of the electric/magnetic fields themselves are held fixed. Applying this limit to the expressions for electro/magneto-static fields of point charges indeed leaves us with constant fields oriented opposite to the direction of the source charges   $\mathbf{\tilde{\Gamma}}$.  Without loss of generality, the z-axis can then be   chosen to point in the direction of the sources. Integrating these fields along the line element, precisely yields the linear potential terms in eq.(\ref{2}).

In fact we may also use this limiting two-center  system  to captures other features of our original configuration of a black hole in constant external    fields.  Following  \cite{D1}, \cite{D2}, \cite{D3}, we can determine the off-diagonal terms in the metric  using 
\begin{equation}
\nabla \times \vec{\omega} =  {\cal P}(\vec{x}) \nabla  {\cal Q}(\vec{x}) - {\cal Q}(\vec{x})  \nabla  {\cal P}(\vec{x}) 
\label{6}
\end{equation} 
Below we shall solve $\vec{\omega}$ for a class of precessing solutions.  Furthermore operating a gradient on both sides of eq.(\ref{6}) leads  to the  following integrability equation 
\begin{equation}
{\cal P}(\vec{x}) \nabla^2  {\cal Q}(\vec{x}) - {\cal Q}(\vec{x})  \nabla^2  {\cal P}(\vec{x})   =   0
\label{7}
\end{equation} 
which we evaluate at $\vec{x} = \vec{l}$ to get 
\begin{equation}
l = \frac{qu - pv}{pE - qB}
\label{8}
\end{equation} 
This gives us the position of the black hole. Here $\vec{l} = (0,0,l)$ can be chosen on grounds of symmetry. One can also perform a shift of coordinates, so as to place the black hole at the origin. This can be achieved by setting constants $u = v = 0$.  Note however that $( pE - qB ) \neq 0$  is required in order to preserve  mutual nonlocality.

Eq.(\ref{6}) can be conveniently solved using spherical coordinates $(r, \theta, \phi)$. And that leads to a system of coupled differential equations
\begin{eqnarray}
\left( \nabla \times \vec{\omega} \right)_{r}  &=&  \frac{2 \; \cos  \theta \, ( pE - qB ) }{r}   \nonumber   \\
\left( \nabla \times \vec{\omega} \right)_{\theta}  &=&  - \; \frac{ \sin \theta \, ( pE - qB ) }{r}
\label{9}
\end{eqnarray} 
while $\left( \nabla \times \vec{\omega} \right)_{\phi}  = 0$ due to $\phi$-independence on the right-hand side.  Our objective is now to seek out a non-trivial solution which confers to the description of a black hole rotating in the presence of external electromagnetic fields. We  find that there exists  such a simple solution with azimuthal symmetry  
\begin{eqnarray}
\omega_{\phi} = \sin \theta \, ( pE - qB )
\label{10}
\end{eqnarray} 
while $\omega_{r} = \omega_{\theta} = 0$.  For completeness let us also mention  that the solution presented in eq.(\ref{10}) is certainly not the most general.  For instance, we also find that solutions with harmonic variations such as $\frac{\partial \omega_{\theta}}{\partial \phi} = \cos \phi$   also exist and very likely one may well find a more general class of these. But we shall not require that for our purposes.

The solution above allows us to levitate a black hole at a fixed height on the $xy$-plane owing to the balancing act between gravitational attraction and electro/magneto-static repulsion. However it is not stable in all three directions and can move about the surface of the plane.  To localise the black hole in  all three directions we need a more complicated background field where the  black hole can be held at a local minimum of an effective potential.

\section{ Continuum Limit of Multi-Center Solutions }
In this section we start looking for extremal stationary solutions to Einstein-Maxwell gravity that admit backgrounds with multipole electromagnetic fields.  As before, we work with four dimensional gravity with just one gauge field.   Generalisations to $n - 1$ vector fields or inclusion of other charges such as D0 and/or D6 in Type II A are rather straightforward.  Let us now see how     taking a continuum limit of Denef et al's multi-center solutions yields the desired backgrounds.  In order to write down harmonic functions for such a  smeared distribution of black holes,  we define density functions $\rho_e(\vec{x}^{\prime})$, $\rho_m(\vec{x}^{\prime})$ via
\begin{equation}
\int_V  \rho_e(\vec{x}^{\prime})  d\tau^{\prime}  =  Q   \qquad  \mbox{and}  \qquad    \int_V  \rho_m(\vec{x}^{\prime})  d\tau^{\prime}  =  P  
\label{3.4} 
\end{equation} 
where $d \tau^{\prime}$ is a volume element within a compact support $V$, that   covers the distribution. In the continuum limit, harmonic functions for multiple black holes  take the form
\begin{equation}
{\cal Q}(\vec{x}) = v + \int_V  \frac{\rho_e(\vec{x}^{\prime})}{|\vec{x} - \vec{x}^{\prime}|} d\tau^{\prime}   \qquad  {\cal P}(\vec{x}) = u +  \int_V  \frac{\rho_m(\vec{x}^{\prime})}{|\vec{x} - \vec{x}^{\prime}|} d\tau^{\prime}   
\label{3.5} 
\end{equation} 
To these harmonics one may also add linear terms $Ez$ and $Bz$ corresponding to  constant fields, whenever required.  From a computational point of view, the real utility of the above-mentioned smeared distributions shows up in their  respective multipole expansions. Expressing this in the regime that  $|\vec{x}| >> |\vec{x}^{\prime}|$ holds, we have
\begin{eqnarray}
{\cal Q}(\vec{x}) = v +  \frac{Q}{|\vec{x}|}  +  \frac{x_i \Delta^i_e}{|\vec{x}|^3}  +  \frac{1}{2} \frac{x_i x_j T_e^{ij}}{|\vec{x}|^5}  +    \cdots \cdots    \nonumber   \\
{\cal P}(\vec{x}) = u +  \frac{P}{|\vec{x}|}  +  \frac{x_i \Delta^i_m}{|\vec{x}|^3}  +  \frac{1}{2} \frac{x_i x_j T_m^{ij}}{|\vec{x}|^5}  +    \cdots \cdots  
\label{3.7} 
\end{eqnarray} 
where $Q$, $P$ are electric respectively magnetic monopole moments; $\bf{\Delta_e}$,    $\bf{\Delta_m}$  are electric and magnetic dipole moment vectors;  and  $\bf{T_e}$,  $\bf{T_m}$  are respectively electric and magnetic quadrupole moment tensors - all  defined in the usual way.  We employ boldface characters to denote vectors as well as tensors. The ``$\cdots \cdots$''  in eq.(\ref{3.7})  denote terms with higher order moments.  When $|\vec{x}| >> |\vec{x}^{\prime}|$, the series is convergent and these  functions can  be used to describe  supergravity solutions associated to any specific multi-moment  source, provided all lower moments vanish for that distribution. As an   illustrative example,  we analyse the solution for a charge distribution with dipole order corrections. 

First let us check that the functions in eq.(\ref{3.5}) yield integrability conditions that are well-defined in the continuum limit. Evaluating  eq.(\ref{7}) for these  harmonics gives
\begin{eqnarray}
\rho_e(\vec{x}) {\cal P}(\vec{x}) - \rho_m(\vec{x}) {\cal Q}(\vec{x})  = 0
\label{3.8} 
\end{eqnarray}
Outside the support $V$, this expression vanishes identically; whereas      points $\vec{x} \in V$ inside the support region yield
\begin{eqnarray}
u \rho_e(\vec{x}) -  v  \rho_m(\vec{x})  = 0
\label{3.9} 
\end{eqnarray}
In order to compare this to the analogous result for a single black hole,  we integrate both sides of eq.(\ref{3.9}) over the volume $V$ to get
\begin{eqnarray}
u Q - v P = 0
\label{3.10} 
\end{eqnarray}
This is exactly what one has for a single-center solution with charges $Q$ and $P$; thereby confirming the asymptotic dependence of $u$ and $v$  for an arbitrary multi-center configuration having fixed total (monopole) charges $Q$ and $P$. 

Having checked consistency of  integrability conditions, we next compute the off-diagonal elements $\vec{\omega}$ in the metric via
\begin{eqnarray}
\nabla \times \vec{\omega} = - {\cal P}(\vec{x})  {\bf E} (\vec{x}) +  {\cal Q}(\vec{x})  {\bf B} (\vec{x})
\label{3.11}
\end{eqnarray}
where ${\bf E} (\vec{x})$ and ${\bf B} (\vec{x})$ refer to exact electric and magnetic fields corresponding to distributions  $\rho_e(\vec{x})$  and  $\rho_m(\vec{x})$  respectively.  In this sense the continuum limit described here is much simpler than a finite $N$ many body black hole system for which integrability equations turn out to be quite hard to solve in full generality.

For our objectives, it will suffice to solve eq.(\ref{3.11}) using its   multipole expansion.  As an illustration, we consider a smeared distribution where the  monopole contributions to  $\vec{\omega}$ get magnetic dipole corrections coming from $\bf{\Delta_m}$, which is aligned along the z-axis. In spherical coordinates,  eq.(\ref{3.11})  takes the form   
\begin{eqnarray}
\left( \nabla \times \vec{\omega} \right)_r &=&   \frac{2 \, v \, \Delta_m \, \cos \theta}{r^3}  +   \frac{Q \, \Delta_m \, \cos \theta}{r^4}  \nonumber     \\
\left( \nabla \times \vec{\omega} \right)_{\theta}  &=&  \frac{v \, \Delta_m \, \sin \theta}{r^3}  +  \frac{Q \, \Delta_m \, \sin \theta}{r^4}  
\label{3.12} 
\end{eqnarray}
while $\left( \nabla \times \vec{\omega} \right)_{\phi} = 0$ due to symmetry in the $\phi$-direction. Note that whilst writing down eq.(\ref{3.12}), we make use of the integrability constraint eq.(\ref{3.10}) ( inserting it into      eq.(\ref{3.11}) ).  As before, we seek solutions characterised by azimuthal symmetry.  The ensuing result is  
\begin{eqnarray}
\omega_{\phi} =  \frac{v  \Delta_m  \sin \theta}{r^2}  +  \frac{Q  \Delta_m \sin \theta}{2 \, r^3}
\label{3.13} 
\end{eqnarray}
and $\omega_{r} = \omega_{\theta} = 0$. At large distances away from the smeared sources, eq.(\ref{3.13}) gives dipole corrections to leading order contributions in the metric. In fact these constitute sub-leading contributions to the geometry.  It is these multipole corrections that distinguish a true one-centered black hole from a multi-center distribution of black holes, when viewed at asymptotic infinity. For a pure one-center solution, $\vec{\omega}$ identically vanishes. While for the multi-center case, it is non-trivial but quite difficult to compute for any given discrete configuration. The continuum limit, on the other hand, facilitates viable computations, at least order by order in a multipole series expansion.

\section{Towards a Black Hole Levitron\texttrademark}
We are now ready to combine  results of the last two sections to construct stable levitating black hole solutions and realise a Levitron$^{TM}$-like construction.  We perturb the constant background fields of section II with a magnetic dipole field and over this perturbed background solve for a black hole held at a fixed height. The dipole fields are produced by the smeared distribution discussed in section III. For simplicity we consider a black hole with only electric charge $q$ ( a dyonic generalisation is also straightforward   ). This construction is captured by the following harmonics
\begin{eqnarray}
\hspace{-0.3cm} {\cal Q}(\vec{x}) = v + \frac{q}{|\vec{x} - \vec{l}|} + Ez  \quad   {\cal P}(\vec{x}) = u + \frac{\Delta_m \cos \theta}{|\vec{x}|^2} + Bz \hspace{0.1cm}
\label{4.1} 
\end{eqnarray}
The dipole moment is aligned parallel to the z-axis  and carries a magnitude  $\Delta_m$.  While  $\theta$ is a coordinate denoting the angle that the position vector $\vec{x}$  makes with the z-axis. However it will suffice to turn off the constant fields $E$ and $B$ for the rest of the computation.  As we shall see this is because a  dipole background  is sufficient to hold the black hole at a fixed height and keep it stable in  all three directions. Superposing constant fields do not affect stability of the solution but ultimately we will need the constant fields for giving an interpretation of black hole levitation in a constant gravitational field ( as would be the case if we were ever to trap a small black hole in a laboratory somewhere on Earth ! ).

Continuing with the calculation, the position of the black hole  $\vec{l}$  is determined by evaluating  eq.(\ref{7}) at the location of the pole $\vec{x} = \vec{l}$ using harmonics in eq.(\ref{4.1}) with $E = B =0$. This gives
\begin{equation}
| \, \vec{l} \, |  \, =  \,  \sqrt{\frac{ -\Delta_m  \cos \theta }{u}}
\label{4.2} 
\end{equation} 
Also evaluating the integrability equation at the other pole $\vec{x} = 0$ determines the constant $v$
\begin{equation}
v = - \frac{q}{|\, \vec{l} \,|}
\end{equation} 
Physical solutions only exist for $l$ ($\equiv  | \, \vec{l} \, |$)  real and non-negative. For instance, when  $u \, > \, 0$, then $\theta$ can attain values from $0$ to $\frac{\pi}{2}$  provided the   dipole is directed along the negative z-axis. $\phi$ remains unconstrained. On the other hand, for a dipole pointing in the positive z-direction, the angle   $\theta$ spans within the range $\frac{\pi}{2}$ to $\pi$. (see fig. 1 below). When $u \, < \, 0$ the signs   appropriately reverse. The solution space of the black hole is now confined to a restricted parameter space. More precisely these are circular orbits corresponding to given values of $\theta$  on an equipotential surface of a dipole field. And in turn each orbit refers to a solution with a specified radial distance $l$. 

In fig. 1 below, we plot eq.(\ref{4.2}).  At $\theta = 0$ the  black hole sits at a fixed height on the z-axis; at $\theta = \frac{\pi}{2}$ it falls into the origin; while the case $0 < \theta < \frac{\pi}{2}$  corresponds to the black hole being located anywhere on a circular orbit  centered at height $l \, \cos  \theta$  and having radius   $l \, \sin  \theta$. Solutions on the positive $z$-axis correspond to the case when $\Delta_m < 0$ (for $u \, > \, 0$), while those on the negative axis refer to $\Delta_m > 0$.  For each value of $\theta$ in  eq.(\ref{4.2}) there exists a solution for $\vec{\omega}$. At $\theta = 0$ the solution space is just a single point and that is when the black hole achieves stability in all three directions at a fixed height on the z-axis. 

For completeness we first compute $\vec{\omega}$ when the black hole is still sitting at the  origin, that is when $\vec{l} = 0$. After that we shall determine the modification in $\vec{\omega}$ required to achieve stable levitation at a fixed height on the z-axis.  In fact the solution at $\vec{l} = 0$. can simply be borrowed from our calculation in  eq.(\ref{3.13}) once we make the substitutions $Q \rightarrow q$ and $P \rightarrow 0$.  

On the other hand, when the black hole is made to levitate at a fixed height $l$ on the z-axis we have to solve the following system of equations  
\begin{widetext}
\begin{eqnarray}
\left( \nabla \times \vec{\omega} \right)_r &=&  -\frac{ q \, u  \left( r - l \, \cos \theta \right) }{ \left( r^2 + l^2 - 2rl \, \cos \theta  \right)^{\frac{3}{2}} } -  \frac{2 \, q \, \Delta_m \, \cos    \theta}{l \, r^3}  - \frac{q \, \Delta_m \, \cos \theta  \left( r  - l \, \cos  \theta \right) }{ r^2  \left( r^2 + l^2 - 2rl \, \cos  \theta  \right)^{\frac{3}{2}} }   +  \frac{2 \, q \, \Delta_m \, \cos  \theta}{r^3   \left( r^2 + l^2 - 2 \, r \, l \, \cos \theta \right)^{\frac{1}{2}} }   \nonumber     \\
\left( \nabla \times \vec{\omega} \right)_{\theta}  &=&   -\frac{ q \, u \, l \, \sin  \theta }{ \left( r^2 + l^2 - 2rl \, \cos  \theta  \right)^{\frac{3}{2}} } -  \frac{ q \, \Delta_m \, \sin  \theta}{l \, r^3}  - \frac{q \, l \, \Delta_m \, \sin  \theta  \, \cos  \theta }{ r^2  \left( r^2 + l^2 - 2rl \, \cos  \theta  \right)^{\frac{3}{2}} }   +  \frac{ q \, \Delta_m \, \sin  \theta}{r^3  \left( r^2 + l^2 - 2 \, r \, l \, \cos \theta \right)^{\frac{1}{2}} }
\label{4.3} 
\end{eqnarray}
\end{widetext}
and again $\left( \nabla \times \vec{\omega} \right)_{\phi} = 0$. Also $\vec{l} = (0, 0, l)$.  This now becomes fairly more complicated compared to the non-levitating case. The modification in the metric reflects a modification to the geometry of the system.  If we restrict to azimuthally symmetric cases,    we find that eq.(\ref{4.3}) has a solution only for small heights of levitation, that is when $l << r$. This can be understood in the following way.  In this set-up the system consists of the black hole plus the source of the dipole field. Let us call the latter the base. The levitating we are looking for requires that the base be rigid against the gravitational pull  of the black hole, that is the center of mass of the whole system be as close to the base as possible.  For very large charges, corresponding to large values of $l$, a stable symmetric levitating solution does not seem to exist ( we see this from  numerical checks ).  In that case more complicated non-symmetric  solutions may be sought for, but we would hardly call those levitating.

Narrowing down to our regime of interest, we expand around  $l << r$ and solve     eq.(\ref{4.3})  order by order in $l$.  Truncating up to second order terms  we get
\begin{widetext}
\begin{eqnarray}
\omega_{\phi} &=&  -\frac{ q \, u  \left( 1 \, - \, \cos  \theta \right)  }{r \, \sin \theta} \; -  \; \frac{q \, \Delta_m \, \sin   \theta}{l \, r^2}  \; + \;  \frac{q \, \Delta_m \, \sin  \theta}{2 \, r^3} \; -  \;  \left\{  \frac{  q \, u  \sin  \theta  }{r^2} \right\} \cdot l  \nonumber \\  &\,& \hspace{0.1cm} + \; \left\{ -\frac{3 q \, u  \cos  \theta \, \sin  \theta  }{2 \, r^3}  \; - \;  \frac{q \, \Delta_m  \left( 1 \, + \, 3 \, \cos^2  \theta   \right) \sin   \theta}{8  \, r^5} \right\} \cdot l^2  \; + \;  {\cal O}(l^3)
\label{4.4} 
\end{eqnarray}
\end{widetext}
while $\omega_{r} = \omega_{\theta} = 0$.  This solution enables us to write down the full metric for a stationary system of a black hole levitating in equilibrium above a magnetic dipole field.  Also this calculation easily extends to the case of a dyonic black hole.

\subsection*{Comparision to a Levitron\texttrademark}
We now compare the levitation of black holes discussed above with that of a $Levitron^{TM}$\cite{WH}. The latter is a spin stabilised magnetic levitation device first invented by Roy Harrigan\cite{RH}.  It basically consists of a permanent base magnet above which a spinning top with a magnetic dipole moment levitates mid-air and is stable in all three directions.  This gives rise to an apparent paradox due to Earnshaw's theorem \cite{E} which states that no  stationary configuration composed of electric/magnetic charges and masses can be held in stable equilibrium purely by static forces. And the reason for this is simply that all static potentials satisfy  Laplace's equation whose solutions only exhibit saddles at critical points : there are neither any maxima nor minima.  It was Sir Michael Berry's \cite{B} (see also \cite{SHR})  remarkable insight invoking adiabatic averaging that helped resolve the apparent paradox. He showed that a slow precession mode ( when averaged over the  
\begin{figure} [ht]
\centering
\includegraphics[totalheight=0.4 \textheight]{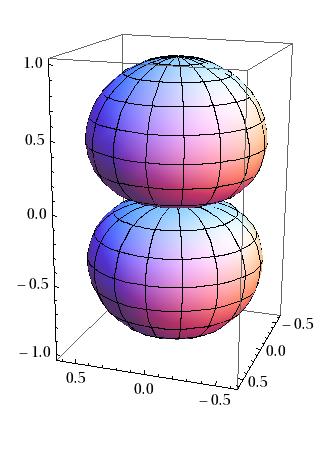}
\caption{ Plotting the solution space for $\vec{l}$ }
\end{figure}
fast rotation mode ) was responsible for creating an effective stationary potential with a stable minimum.  This is the same principle used in neutron traps as well as other particles carrying magnetic dipole moment.

A natural question which arises is whether our black hole construction also mimics the   physics of the $Levitron^{TM}$ and how it overcomes Earnshaw's theorem.  The latter it already seems to evade since it is based on Einstein's gravity rather than Newton's. However the gravitational interpretation of our Black Hole   $Levitron's^{TM}$  balancing mechanism admittedly requires further investigation. Nevertheless a naive classical intuition can be obtained from the fact that a non-vanishing Poynting vector gives rise to a rotating black hole geometry and in turn a rotating electric distribution induces a magnetic field that repels the base magnet. It is the $\vec{\omega}$ in the metric that is responsible for inducing this balancing force. On the other hand the gauge theoretic interpretation of this multi-center balancing has been better   understood in terms of Denef's quiver quantum mechanics \cite{D4}  wherein the distance between centers is determined via an effective potential whose minima determine the stability loci $\vec{l}$.

\section{Conclusions and discussion}
In this article we have constructed a levitating black hole solution. Our Black Hole $Levitron^{TM}$ stabilises an extremal black hole at a fixed location in  an electromagnetic field produced by a continuous   distribution. Our work is built-up using Denef et al's multi-center solutions, which by themselves are stable, stationary BPS solutions with non-local charges.  Our harmonic functions and integrability conditions can all be retrieved as special limits of the discrete multi-center case. Therefore our levitating solutions also describe stable, stationary configurations. This black hole construction very much resembles a mechanical $Levitron^{TM}$ and it would be interesting to investigate if Berry's mechanism can be proven to apply to this set-up as well.  And finally it would be of practical relevance ( in future ! ) to construct solutions  for non-extremal  Black Hole $Levitrons^{TM}$ !

Other interesting directions might be further investigation into other applications of the continuum limit of multi-center solutions. Compared to  discrete-centered configurations, the smeared distribution lends itself to more viable computations. One may ask  what role these distributions play in microstate counting of multiple-black hole geometries. 

\bigskip

\centerline{\bf Acknowledgments}
It is a pleasure to thank P. McFadden, I. Messamah, K. Papadodimas, T. Quella, B. van Rees and M. Shigemori for valuable discussions. This work has been supported in part by the Foundation of Fundamental Research on Matter (FOM).

\end{document}